\documentclass{article}
\usepackage{amsmath}
\usepackage{aip}

\input{tcilatex}

\begin{document}

\title{Correlated weak bonds as a source of the Boson peak in glasses}
\author{Z. Shemer and V. Halpern* \\
Department of Physics, Bar-Ilan UIniversity, \\
Ramat-Gan 52900, Israel.}
\date{\today }
\maketitle

\begin{abstract}
Many attempts to explain the Boson peak in the vibrational spectra of
glasses consider models of a lattice of harmonic oscillators connected by
spring constants of varying strength and randomly distributed. However, in
real glasses one expects that some molecules will be connected to their
neighbors by more than one weak bond, so that a realistic model should
consider oscillators with several weak springs. In this paper, a t-matrix
formalism is used to study the effect of such correlated weak springs in a
scalar model on a simple cubic lattice with a binary distribution of spring
constants. Our results, which are confirmed by computer simulations, show
that a concentration of $c$ oscillators with $z$ weak springs and $6-z$
strong ones leads to a low frequency peak in the reduced density of states
(Boson peak) even when the total concentration of weak springs $cz$ is less
than 10\%., No such peak has been found at these low concentrations in
previously reported calculations which used effective medium methods. For a
given value of $cz,$ this peak becomes more pronounced and moves to much
lower frequencies as $z$ increases.
\end{abstract}

\section{Introduction}

One of the outstanding problems in the theory of glasses is the origin of
the Boson peak. i.e. the low-frequency peak in the reduced density of
vibrational states $g(\omega )/\omega ^{2}$, where $g(\omega )$ is the
density of states at frequency $\omega $. This peak corresponds to an excess
density of states over the usual Debye density, which is proportional to $%
\omega ^{2}$. One of the simplest proposals about its origin is that it
arises from a distribution of the bond strengths between neighboring atoms
in the glass. To check whether this could be the case, numerous studies have
been carried out of the vibrational spectra of sets of harmonic oscillators
on a lattice with different distributions of the oscillator strengths or
spring constants, including a binary distribution \cite{Schir} \cite{Bunde},
Gaussian distributions (including some with negative spring constants) \cite%
{Schir2}, and an inverse frequency distribution\cite{Kant} \cite{Bunde}. All
of these studies considered random (uncorrelated) distributions of spring
constants, for which the density of vibrational states can be calculated by
the coherent medium approximation (CMA), which is equivalent to the
well-known and very powerful coherent potential approximation (CPA) \cite%
{CPA}. However, it is quite possible that in real glasses some atoms are
connected to their neighbors by more than one weak bond. For instance, in a
locally strained region the distances between the atoms may be larger than
the average, so that an atom in the center of such a region is connected to
some or all of its neighbors by weaker bonds. Other possible cases are
interstitial atoms, which can be present even in glasses \cite{Granato}, and
regions of soft potential \cite{soft-pot}. Such correlations have also been
found in a computer generated model of amorphous silicon \cite{Chris} Thus,
it is very interesting to attempt to study the properties of systems of
oscillators in which some of the oscillators have several weak spring
constants. i.e. of correlated weak springs.

This problem of correlated disorder is a very difficult one. For electronic
systems, the homomorphic coherent cluster approximation (HCPA) \cite{HCCPA}
was proposed and had some success. However, in addition to the practical
difficulties of using it there for clusters of more than two sites, a
fundamental objection has been raised to its application for such systems %
\cite{HCCPA-obj}. In our case, the corresponding objection is that the same
oscillator is treated as belonging both to the cluster and to the
surrounding medium. In this paper, we adopt a different approach and
consider low densities of\ oscillators having several weak spring constants,
i.e. clusters of correlated weak springs. The scattering of vibrations from
a single defect can be calculated by the t-matrix method \cite{Econ}, and we
assume that the defects are sufficiently far apart for the scattering from
each defect to be treated independently. In principle, one might think that
the method could be extended to higher defect densities by the use of the
average t-matrix approximation (ATA) \cite{ATA}, which is simpler to use
than the CPA and often gives very similar results for the density of states.
Superficially this seems quite attractive, but it involves theoretical
problems similar to those of the HCPA and grave computational difficulties.
Instead, since the density of states involves only the trace of the Green's
function, we introduce the average trace approximation (AVTA), in which the
trace of the t-matrix of the system is replaced by the sum of the traces of
the t-matrices for each defect. The justification for this is contained in
the Appendix, while the accuracy of the AVTA is demonstrated in section 3 \
Such clusters have also been considered by Gonis and Gorland \cite{Gonis} in
their embedded cluster method, which is an effective medium method and so
spreads out the effects of individual clusters.

The model system that we consider is a simple cubic lattice of coupled
harmonic oscillators with spring constants $K$. For the sake of simplicity
we chose a binary distribution, with a small fraction of oscillators having
a spring constant $K_{1}$ (which we refer to as weak springs) much less than
the spring constant $K_{0}$ of most of the oscillators. Extensive results
for a random distribution of weak springs in this system have been presented
by Schirmacher and Diezemann.\cite{Schir}, who found that a concentration of
10\% of weak springs had a negligible effect on the density of vibrational
states. In contrast to their model, we consider systems in which the weak
springs occur in clusters, so that there are defects in which a site is
linked to its six neighbors by $z$ weak springs with spring constant $K_{1}$
and $6-z$ ordinary springs with spring constant $K_{0}$. The concentration $%
c $ of these defects was chosen to be such that the total fraction $zc$ of
weak springs was less than 10\%, and most of our calculations were performed
for $K_{1}/K_{0}=0.1$. This system and the AVTA method for calculating the
t-matrix and density of vibrational states are described in Section 2 and
the Appendix, while the results of our calculations are presented in Section
3. Even for $z=2$ there were signs of a small Boson peak close to the first
van-Hove singularity, and for $z=6$ a strong Boson peak was found at a much
lower frequency. We also present the results of computer simulations \ for
lattices of up to $16^{3}$ sites which agree well with our calculations. In
section 4 we discuss the origins of the extra low-frequency vibrations that
we found, and our conclusions about the importance of correlations between
weak bonds, as well as about the efficiency of the AVTA, are presented in
section 5.

\section{The model system and the average trace approximation (AVTA)}

The model system that we consider is that of the scalar vibrations of a
simple cubic lattice of coupled harmonic oscillators with spring constants $%
K $. For the sake of simplicity we chose a binary distribution of spring
constants, with a small fraction of oscillators having a spring constant $%
K_{1}$ (which we refer to as weak springs) much less than the spring
constant $K_{0}$ of most of the oscillators. We write the Hamiltonian for
this system in the form 
\begin{equation}
H=H_{0}+H_{1}  \tag{1}
\end{equation}%
where $H_{0}$ is the Hamiltonian of a system without any weak springs, 
\begin{equation}
H_{0}=\epsilon _{0}\sum_{\mathbf{n}}|\mathbf{n}><\mathbf{n}|+%
{\frac12}%
K_{0}\sum_{\mathbf{n}}\sum_{\mathbf{a}}\{|\mathbf{n}><\mathbf{n}|-|\mathbf{n}%
><\mathbf{n+a}|\},  \tag{2}
\end{equation}%
and $H_{1}$ is the Hamiltonian of the ''perturbation'' due to the weak
springs connecting sites $\mathbf{m}$\textbf{\ }and\textbf{\ }$\mathbf{m}+%
\mathbf{a}$,%
\begin{equation}
H_{1}=%
{\frac12}%
(K_{1}-K_{0})\sum_{\mathbf{m}}\{|\mathbf{m}><\mathbf{m}|+|\mathbf{m+a}><%
\mathbf{m+a}|-|\mathbf{m}><\mathbf{m+a}|-|\mathbf{m+a}><\mathbf{m}|\} 
\tag{3}
\end{equation}%
We consider a system in which the weak springs are arranged in disjoint
clusters. As explained in the Appendix, for each cluster $\alpha $ we define
a vector $|\alpha >=(|\mathbf{n}_{\alpha }\mathbf{>,}|\mathbf{n}_{\alpha }%
\mathbf{+a}_{1}\mathbf{>,}|\mathbf{n}_{\alpha }\mathbf{+a}_{2}\mathbf{>,....)%
}$ containing all the sites belonging to that cluster, and write 
\begin{equation}
H_{1}=\sum_{\alpha =1}^{M}H_{\alpha }=\sum_{\alpha =1}^{M}|\alpha >V_{\alpha
}<\alpha |,  \tag{4}
\end{equation}%
where $V_{\alpha }$ is the square matrix%
\begin{equation}
V_{\alpha }=<\alpha |H_{1}|\alpha >.  \tag{5}
\end{equation}%
We define in a similar way the matrix $G_{\alpha }=<\alpha |G_{0}|\alpha >$,
where $G_{0}$ is the Green function of the Hamiltonian $H_{0}$. It then
follows that the t-matrix $t_{\alpha }$ corresponding to the isolated
cluster $|\alpha >$ is 
\begin{equation}
t_{\alpha }=V_{\alpha }[I-G_{\alpha }V_{\alpha }]^{-1}  \tag{6}
\end{equation}%
and the corresponding operator $T_{\alpha }$ is%
\begin{equation}
T_{\alpha }=|\alpha >t_{\alpha }<\alpha |=\sum_{k,l}t_{kl}|\mathbf{m}_{k}><%
\mathbf{m}_{l}|.  \tag{7}
\end{equation}%
where $\mathbf{m}_{k}=\mathbf{n}_{\alpha }\mathbf{+a}_{k}$. In order to
calculate the density of states for a set of $M$ identical, randomly placed,
disjoint clusters, we use the \textbf{av}erage \textbf{t}race \textbf{a}%
pproximation (AVTA), in which we write

\begin{equation}
\langle Tr\{G_{0}TG_{0}\}\rangle \approx \langle Tr\{G_{0}(\sum_{\alpha
=1}^{M}T_{\alpha })G_{0}\}\rangle =M\langle Tr\{G_{0}|\alpha >t_{\alpha
}<\alpha |G_{0}\}\rangle ,  \tag{8}
\end{equation}%
where $\langle .\rangle $ denotes the mean value. For the calculation of $%
Tr\{G_{0}|T_{\alpha }|G_{0}\}$ for a given cluster $\alpha $, we write $%
G_{0}(z)$ in the form $G_{0}(z)=\sum_{\mathbf{k}}|\mathbf{k}><\mathbf{k}%
|/(z-E_{\mathbf{k}})$ and use equation (7) for $T_{\alpha }$, while $<%
\mathbf{k|n}>=\exp (-i\mathbf{k}\cdot \mathbf{n})$. It then follows that

\begin{equation}
Tr\{G_{0}(z)|T(z)|G_{0}(z)\}=\frac{M}{N}\sum_{l,m}\sum_{\mathbf{k}}t_{lm}%
\frac{\exp [i\mathbf{k}\cdot (\mathbf{a}_{m}-\mathbf{a}_{l})]}{(z-E_{\mathbf{%
k}})^{2}},  \tag{9}
\end{equation}%
which depends only on the distances $|\mathbf{a}_{m}-\mathbf{a}_{l}|$
between pairs of sites in a cluster. In order to evaluate the trace in terms
of the known expressions for $G_{0}(z;\mathbf{a}_{l},\mathbf{a}_{m})$, we
write equation (9) in the form

\begin{equation}
Tr\{G_{0}(z)|T(z)|G_{0}(z)\}=-\frac{M}{N}\frac{\partial }{\partial z}%
\sum_{l,m}t_{lm}G_{0}(z;\mathbf{a}_{l},\mathbf{a}_{m}).  \tag{10}
\end{equation}%
Since in general the density of states is%
\begin{equation}
g(E)=\mp \frac{1}{\pi }\func{Im}\{Tr[G(E\pm i\epsilon )]\},  \tag{11}
\end{equation}%
while $G=G_{0}+G_{0}TG_{0}$, it follows that the density of states for our
system in the AVTA is 
\begin{equation}
g(E)=g_{0}(E)+\frac{M}{N}\frac{1}{\pi }\func{Im}\{\frac{\partial }{\partial z%
}\sum_{l,m}t_{lm}G_{0}(E-i\epsilon ;\mathbf{a}_{l},\mathbf{a}_{m})\}. 
\tag{12}
\end{equation}

\section{Results of our calculations.}

We performed calculations and obtained similar results for various low
concentrations of weak bonds, and report here those for systems with a total
concentration of 9\% of weak bonds. These were distributed in the different
systems as a random concentration of 1.5\% of clusters of 6 weak bonds, of
2.25\% of clusters of 4 weak bonds, of 4.5\% of clusters of 2 weak bonds and
of 9\% of single weak bonds. In addition to calculations by the AVTA, we
also performed numerical simulations for the different systems on cubes with
13-16 oscillators on each side, and took a smoothed average of the
cumulative density of states obtained for these systems. From this, the
density of states $g(\omega )$ was obtained by numerical differentiation.

In order to obtain an estimate of the unit of time in our calculations and
simulation, we note that in our calculations we chose the lattice parameter $%
a=1$ and the average sound velocity as $c_{s}=1$. Since typical values of
these quantities in real materials are $a\symbol{126}10^{-10}m$ and $c_{s}%
\symbol{126}10^{3}m/s$, the unit of time $a/c_{s}\symbol{126}10^{-13}s$.

In order to demonstrate the accuracy of the AVTA, we show in figure 1 the
reduced density of states $g(\omega )/\omega ^{2}$ as a function of the
frequency $\omega $ for clusters of 6 weak bonds as found by the AVTA and by
our numerical simulations for $K_{1}/K_{0}=0.1$, and also its value for the
system without any weak bonds for the sake of comparison. We note first that
in the periodic system the density of states $g(\omega )/\omega ^{2}$ is
fairly smooth, with the well-known van-Hove singularities, where there is a
discontinuity in the gradient of $g(\omega )$, at $\omega =2$ and $\omega =%
\sqrt{8}.$ On the other hand, both the calculations and the simulations show
a large peak in $g(\omega )/\omega ^{2}$ around $\omega =0.75$. The
difference in the heights of the maxima between the calculations and
simulations are associated with the scatter that occurs in the simulations,
and the areas under the peaks, which correspond to the total extra density
of states in this region, are very similar. The slight difference in the
positions of the peak may also be associated with this scatter. However, it
might just be a genuine effect arising from interactions between the defect
clusters which are ignored in the AVTA but could occur in the simulated
systems (where noise places a lower limit to the defect concentrations that
can be studied). Incidentally, in the calculated density of states, there
are slight problems at the frequencies of the Van-Hove singularities,
because of our use of equation (1\textbf{2}), since the derivative of $%
G_{0}(z)$ is discontinuous at these points.

The above problems with the numerical simulations, including the scatter,
arise mainly from our considering the density of states $g(\omega )$ rather
than the accumulated density of states $N(\omega )$. Accordingly, we show in
figure 2 the difference $\Delta N(\omega )$ between the values of $N(\omega
) $ for a system with clusters of six weak springs with $K_{1}/K_{0}=0.1$
and the system without any defects. In this figure, the points are the
results of the simulations, and we see clearly how these are scattered
around the continuous line which shows the results of our calculations. The
dotted line, which shows gtg($\omega )$ for this system, is included in
order to show the position of the peak in the density of states ($g(\omega
)=dN/d\omega $) and the van-Hove singularities for this system \ The
increased scatter for frequencies above the first van-Hove singularity at $%
\omega =2$ is associated with the large number of states contributing to $%
g(\omega )$ in this region, and is of no fundamental significance.

Now that we have established the accuracy of the AVTA method, all the other
results that we present were calculated using this method. In order to show
the effects of different cluster sizes, we present in figure 3 the reduced
density of states $g(\omega )/\omega ^{2}$ as a function of $\omega $, with $%
K_{1}/K_{0}=0.1$, for all the systems that we considered. It is immediately
apparent that as the number of weak bonds in a cluster is reduced from 6 to
4 to2 the position of the peak moves to higher frequencies, while for
randomly placed weak bonds there is no peak (just as was found in \cite%
{Schir} for a concentration of 10\% weak bonds). In order to throw some
light on the origin of these peaks, as discussed in the next section, we
show in figures 4a and 4b as a function of $\omega $\textbf{\ }the
difference $gtg(\omega )$ in the density of states between the systems with
weak springs and the pure system with no weak springs for the different
sizes of clusters. Since the presence of weak springs cannot affect the
total number of states, the integral of this difference $\int_{0}^{\infty
}gtg(\omega )d\omega $ must vanish, and for all the systems we found that it
was less than $10^{-3}$. Finally, we present in figure 5 the density of
states $g(\omega )$ for clusters of 6 weak bonds with various values of the
ratio $K_{1}/K_{0}$. We show $g(\omega )$ rather than $g(\omega )/\omega
^{2} $ because there is a simple physical meaning to the position of the
peak in the density of states, as discussed in the next section. We just
note for now that as $K_{1}/K_{0}$ increases the peak moves to higher
frequencies, until for $K_{1}/K_{0}=1$, the pure system, it coincides with
the first van-Hove singularity.

\section{Discussion}

The main result presented above is that correlation between weak bonds has a
strong effect on the density of states and leads to the appearance of a
maximum in the reduced density of states, which we will call a boson peak,
as can be seen from figures 1 and 3. In order to understand our results, we
consider first the variation of the position $\omega _{\max }$ of the
maximum in $g(\omega )$ for clusters of 6 weak bonds having different values
of $K_{1}/K_{0}$, as shown in figure 5. The values of $\omega _{\max
}(K_{1}/K_{0})$ were $\omega _{\max }(0.1)=0.769$, $\omega _{\max
}(0.2)=1.075$ and $\omega _{\max }(0.3)=1.309,$ which all satisfy $\omega
_{\max }(K_{1}/K_{0})\propto \sqrt{K_{1}/K_{0}}$. Such a relationship
corresponds to that for the natural resonance frequencies of oscillators
having spring constants $K_{1}$, which suggests that the extra modes are
associated with oscillations localized mainly within these clusters.
However, the degeneracy of these resonant modes with those of the
unperturbed system leads to a spreading in the range of frequencies, and so
to a broadening of the peak, and to less localization of the extra
vibrations. Incidentally, it follows from these values that the resonance
frequency for oscillators with spring constant $K_{0}$ is $\omega _{\max
}(1)=2.43$, which lies between the two van-Hove singularities of the pure
lattice. This explains why the peaks become broader as $K_{1}/K_{0}$
increases, since the resonance frequencies of the clusters and the pure
system become closer together.

The variation of the position of the boson peak with the size of the
clusters of weak bonds shown in figure 3 can be explained in a similar
manner. As the cluster size increases, the peak is expected to move closer
to the eigenfrequency associated with the weak bonds. In order to understand
this effect in more detail, we must first consider the limitations of the
scalar model on a simple cubic lattice, in which the vibrations along the
different axes are independent of each other. Physically, one might expect
that an oscillator between a pair of colinear weak bonds, with the other
four bonds restraining the oscillator strongly along their axes, would give
rise to a localized vibration along them with a frequency close to the
eigenfrequency of the weak bonds. However in the scalar model the motion is
calculated without any reference to the direction. As a result, the
calculated frequency is a sort of average of the frequencies for motion
along the different axes. Accordingly, one pair of weak bonds will
contribute to the formation of a peak at a much higher frequency than that
of three pairs (six weak bonds). Only for a cluster of six weak bonds will
the natural frequency coincide with the eigenfequency of the weak bonds. In
addition, for smaller clusters of weak bonds the strong bonds will lead to a
greater mixing with the vibrations of the lattice of strong bonds, and so to
a greater broadening of the peak.

Since the clusters of weak bonds do not affect the total number of states in
the system, the excess states at low frequencies must be associated with a
reduction in the number of states at high frequencies. Our calculations of $%
gtg(\omega )$ show that for all the systems the excess states are below the
first van-Hove singularity, which we refer to as the low frequency region,
and that these come at the expense of states between the two van-Hove
frequencies (the middle frequency range) and above the second van-Hove
frequency (the high frequency range). Our calculations show that the total
number of states that move to the low frequency range depends solely on the
concentration of the weak bonds, and not on their distribution. The
correlation only affects the position of the peak in the low frequency range
and their origin in the middle and high frequency regions. As can be seen
clearly from figure 4b, as the correlation between the weak bonds increases
so does the number of states that move out of the middle frequency region,
which contains the previously calculated resonance frequency of the pure
lattice $\omega _{0}=2.43$ \ We note that the largest number of states leave
this region for clusters of four and six weak bonds, for which the results
are very similar, and less for clusters of two weak bonds. The smallest
number of states that leave the middle frequency region, and hence the
largest number that leave the high frequency region, is for isolated weak
bonds. The reason for the similarity between the results for clusters of
four weak bonds and those for clusters of six weak bonds is that the former
only contain strong bonds along one axis, and the same sort of averaging
process as mentioned above occurs in the scalar model.

Finally, we consider briefly the relationship between our results and those
of the soft potential model \cite{soft-pot}. The basic concept of this model
is that there are two types of motion and related excitations, namely
acoustic and soft-mode types, which coexist and interact with each other and
so contribute to the boson peak. Our system of a binary distribution of
spring constants with clusters of weak springs is perhaps the simplest
example of such soft potentials, and we found in it from first principles a
low frequency peak in the phonon density of states.

\section{Conclusions}

The most important result of our calculations on a scalar model for a simple
cubic lattice of oscillators show that clustering of weak springs has a
strong effect on the density of states. Even with a concentration of only
9\% of weak springs with spring constants a fifth or a tenth of that of the
pure system, we found that this clustering leads to a very noticeable peak
in the low-frequency reduced density of states. Such effects were not found
in previous studies because these described the system in terms of an
average medium, which cannot take fully into account short-range order. This
peak is very reminiscent of the boson peak universally found in glasses, and
suggests that such correlated defects (similar to soft potentials) may be
one of the causes of this peak.

Another conclusion from our results is that the average trace approximation
is very accurate for the diagonal elements of a system's Green function at
low impurity concentrations, and in principle can be applied to systems with
various types of impurity clusters. In addition, our AVTA method enables us
to identify the source of these extra low frequency states. We found that
for a given concentration of weak springs the same total number of states
move from the higher frequency regions above the first van-Hove singularity
of the pure lattice to the low frequency region below it. However, for
isolated weak springs these extra states all lie so close to this
singularity that they do not produce a peak in the density of states. As the
number of weak springs in a cluster increases, they move to a lower
frequency and become more concentrated around the resulting peak frequency,
and as a result the strength of this peak increases. Our approach can
readily be extended to treat mixtures of clusters of weak springs of
different sizes and with different spring constants, and in principle can
also be applied to other systems.

\section{Appendix - The Average Trace Approximation (AVTA)}

In this appendix, we describe in more detail the use of t-matrices for
impurities associated with clusters of sites, discuss and prove the accuracy
of the AVTA used in equation (8), and discuss its relationship to the
embedded cluster method.

The systems that we consider have a Hamiltonian of the form 
\begin{equation}
H=H_{0}+H_{1}  \tag{A1}
\end{equation}%
where $H_{0}$ is a Hamiltonian having a known Green function $G_{0}$ and the
perturbation (not necessarily small) $H_{1}$ can be expressed in the
appropriate basis set as the sum of Hamiltonians $H_{\alpha },\quad \alpha
=1...M$ associated with $M$ disjoint clusters $\ $For convenience, we will
consider the most common case, in which $H_{0}$ is the Hamiltonian of a
crystal and we use the site representation, in which we denote a site by $%
\mathbf{n}$, so \ that an operator $F$ can be written as $\sum_{\mathbf{m}%
}\sum_{\mathbf{n}}|\mathbf{m}>F(\mathbf{m},\mathbf{n})<\mathbf{n}|$. Our aim
is to calculate the density of states $g(E)$ in the system from the trace of
its Green function \cite{Econ}, 
\begin{equation}
g(E)=\lim_{\epsilon \rightarrow 0+}\left[ \mp \frac{1}{\pi }\func{Im}%
\{Tr[G(E\pm i\epsilon )]\}\right] \text{.}  \tag{A2}
\end{equation}%
For the calculation of $G$, we use the T-matrix formalism, and write 
\begin{equation}
G(z)=G_{0}(z)+G_{0}(z)T(z)G_{0}(z),  \tag{A3}
\end{equation}%
where 
\begin{equation}
T(z)=H_{1}[I-G_{0}(z)H_{1}]^{-1}.  \tag{A4}
\end{equation}%
For random disorder, the above formalism is usually used either in the
average T-matrix approximation (ATA), where one writes $%
<G(z)>=G_{0}(z)+G_{0}(z)<T(z)>G_{0}(z)$,$\ $or in the CPA, in which one
finds the value of $z$ for which $<T(z)>=0$. However, even the extension of
the ATA to clusters of more than two sites involves all the problems of the
HCPA mentioned in the Introduction. In addition, even for two sites there
are problems in defining the operator $P$ required to solve equation (A7),
as described below after equation (A12). The AVTA is an entirely different
method for calculating the diagonal elements of $G$, and hence the density
of states. Our method has some similarity in its method of calculation to
the embedded cluster method of Gonis and Garland \cite{Gonis}, but is based
on an entirely different approach and formalism, as is discussed below.%
\textbf{\ }

While a general operator $F$ has non-zero matrix elements between any pair
of sites, the Hamiltonian $H_{p}$ of the perturbation is the sum of
Hamiltonians $H_{\alpha }$ each of which is associated with a cluster
containing only a few sites. In that case, if a cluster centred around $%
\mathbf{n}_{\alpha }$ contains only the site $\mathbf{n}_{\alpha }$ and some
adjacent sites $\mathbf{n}_{\alpha }+\mathbf{a}_{k},\quad k=1..R$, it is
convenient to introduce the row vector $|\alpha >=(|\mathbf{n}_{\alpha }>,|%
\mathbf{n}_{\alpha }\mathbf{+a}_{1}>,...,|\mathbf{n}_{\alpha }\mathbf{+a}%
_{k}>)$, as proposed by Economou \cite{Econ}. In order to show the
connection between our method and the usual ATA, we consider first the
conventional case where each cluster consists of only a single site, so that
two clusters $\alpha =1,2$ contain only $|\mathbf{n}_{1}>$ and $|\mathbf{n}%
_{2}>$ respectively, and write $G(z;\mathbf{n}_{1},\mathbf{n}_{2}\mathbf{)}$
= $<\mathbf{n}_{1}|G(z)|\mathbf{n}_{2}>$, and similarly $T(z;\mathbf{n}_{1},%
\mathbf{n}_{2}\mathbf{)}=<\mathbf{n}_{1}|T(z)|\mathbf{n}_{2}>$. In this
notation, we can write 
\begin{equation}
H_{1}=\sum_{\alpha =1}^{M}H_{\alpha }=\sum_{\alpha =1}^{M}|\mathbf{n}%
_{\alpha }>V_{\alpha }<\mathbf{n}_{\alpha }|.  \tag{A5}
\end{equation}%
where $V_{\alpha }$ is the matrix (in this case a scalar) describing the
perturbation at site $\mathbf{n}_{\alpha }$. For a system containing only a
single defect at site $\mathbf{n}_{\alpha }$, it readily follows from
equations (A4) and (A5) that the corresponding operator $T_{\alpha }$ is
just $T_{\alpha }=|\mathbf{n}_{\alpha }>t_{\alpha }<\mathbf{n}_{\alpha }|$,
where $t_{\alpha }$ is a scalar, 
\begin{equation}
t_{\alpha }=V_{\alpha }[1-<\mathbf{n}_{\alpha }|G_{0}|\mathbf{n}_{\alpha
}>V_{\alpha }]^{-1}.  \tag{A6}
\end{equation}%
If there are a number of defects, one can write 
\begin{equation}
T=\underset{\alpha }{\tsum }Q_{\alpha }=\underset{\alpha }{\tsum }[T_{\alpha
}+T_{\alpha }G_{0}\underset{\beta \neq \alpha }{\tsum }Q_{\beta }],  \tag{A7}
\end{equation}%
where $Q_{\alpha }$\ is the contribution to the scattering matrix $T$ from
the defect at $\mathbf{n}_{\alpha }$ in the presence of all the other
defects. Iteration of this equation leads to the standard equation 
\begin{equation}
Q_{\alpha }=\underset{\beta \neq \alpha }{\tsum }\,\underset{\gamma \neq
\beta }{\tsum {\tiny \cdot \cdot }}T_{\alpha }+T_{\alpha }G_{0}T_{\beta
}+T_{\alpha }G_{0}T_{\beta }G_{0}T_{\gamma }+...  \tag{A8}
\end{equation}

These equations can be generalized to defects containing more than one site
by replacing the scalars $|\mathbf{n}_{\alpha }>$ with the vectors $|\alpha
> $\ throughout, in which case $V_{\alpha }$, $G_{\alpha }=<\alpha
|G_{0}|\alpha > $ and $t_{\alpha }$ will be square matrices. In order to
show this, we consider a specific cluster $\alpha $, and write $\mathbf{m}%
_{k}=\mathbf{n}_{\alpha }+\mathbf{a}_{k}$. Then, for instance 
\begin{equation}
H_{\alpha }=|\alpha >V_{\alpha }<\alpha |=\sum_{k,l}|\mathbf{m}_{k}>V_{kl}<%
\mathbf{m}_{l}|,  \tag{A9}
\end{equation}%
and since $<\mathbf{r}|\mathbf{s}>=\delta _{\mathbf{r},\mathbf{s}}$ it
follows that $<\mathbf{r|}H_{\alpha }|\mathbf{s}>=\delta _{\mathbf{r},%
\mathbf{m}_{k}}\delta _{\mathbf{s},\mathbf{m}_{l}}V_{kl}$, and so that $%
V_{kl}=<\mathbf{m}_{k}|H_{\alpha }|\mathbf{m}_{l}>$ is the element $(k,l)$
of the matrix $V_{\alpha }=<\alpha |H_{\alpha }|\alpha >$. Similar
definitions apply to the matrices $G_{\alpha }=<\alpha |G_{0}|\alpha >$ and $%
t_{\alpha }$, and it can easily be shown that the equivalent of equation
(A6) for our matrices is 
\begin{equation}
t_{\alpha }=V_{\alpha }[I-G_{\alpha }V_{\alpha }]^{-1}  \tag{A10}
\end{equation}%
Thus, in general we can write 
\begin{equation}
T=\sum_{\alpha }\underset{\beta \neq \alpha }{\tsum }\underset{\gamma \neq
\beta }{\tsum {\tiny \cdot \cdot }}T_{\alpha }+T_{\alpha }G_{0}T_{\beta
}+T_{\alpha }G_{0}T_{\beta }G_{0}T_{\gamma }+...,  \tag{A11}
\end{equation}%
where now 
\begin{equation}
T_{\alpha }=|\alpha >t_{\alpha }<\alpha |=\sum_{k,l}t_{kl}|\mathbf{m}_{k}><%
\mathbf{m}_{l}|.  \tag{A12}
\end{equation}%
However, while for single site clusters equation (A7) can be solved by
defining an operator $P=G_{0}-<\mathbf{n}_{\alpha }|G_{0}|\mathbf{n}_{\alpha
}>$, this method does not apply to solving equation (A11) for clusters
containing more than one site. Instead, we note that for the density of
states we only need the trace of $G$ and so of $G_{0}$ and of $G_{0}TG_{0}$.
In order to calculate the latter, we introduce the average trace
approximation (AVTA) mentioned in section 2. In this approximation we write 
\begin{equation}
\langle Tr\{G_{0}TG_{0}\}\rangle \approx \langle Tr\{G_{0}(\sum_{\alpha
=1}^{M}T_{\alpha })G_{0}\}\rangle =M\langle Tr\{G_{0}|\alpha >t_{\alpha
}<\alpha |G_{0}\}\rangle ,  \tag{A13}
\end{equation}%
where $\langle .\rangle $ denotes the mean value and $M$ is the number of
defect clusters. This is just equation (8) of the paper. In the AVTA it is
essential to take the trace before averaging, since we do not average out
the defects over the whole system and calculate $\langle G\rangle $ for the
periodic average system as is done in the ATA and CPA, as well as in the
embedded cluster method of Gonis and Gorland \cite{Gonis} .

The justification of the AVTA for small defect densities is that the first
term which we are ignoring in $Tr\{G_{0}TG_{0}\}$ is $Tr\{G_{0}T_{\alpha
}G_{0}T_{\beta }G_{0}\}$ where $\beta \neq \alpha $. On taking the trace in
the site representation, we write 
\begin{equation}
Tr\{G_{0}T_{\alpha }G_{0}T_{\beta }G_{0}\}=\sum_{\mathbf{n}}<\mathbf{n}%
|G_{0}|\alpha >t_{\alpha }<\alpha |G_{0}|\beta >t_{\beta }<\beta |G_{o}|%
\mathbf{n}>.  \tag{A14}
\end{equation}%
Each term of this sum involves the matrix elements of $G_{0}$ between sites
belonging to two different defects multiplied by the matrix elements of $%
G_{0}$ between a site $\mathbf{n}$ and one site from each defect. These
terms will be small if the defects are far apart, as they will be for low
defect densities. In terms of Feynman diagrams, the first terms that we are
ignoring correspond to propagation from an arbitrary site to one defect,
from that to the second defect, and then back to the original site. If we
were interested in non-diagonal elements of $G$, for some of which a path
starts at a site close to cluster $\alpha $ and ends at a site close to
cluster $\beta $, this approximation would not be justified. Equation (A11)
is formally similar to equation (2.11) of Gonis and Gorland \cite{Gonis} in
their embedded cluster method. However, their whole approach is based on
finding the self-energy of an effective medium containing the cluster, which
spreads out the effects of the clusters. One major disadvantage of their
method is that, in contrast to the AVTA, it does not distinguish between the
contributions of the pure lattice and the clusters of weak springs. Our
approach, on the other hand, is to use the t-matrix to calculate explicitly
the changes produced by clusters to the original medium that contains no
defects, as shown in the paper. This is especially important for systems
with low densities of defect clusters.

\newpage

Captions for figures.

Fig.1 (Color on line) The reduced density of states $g(\omega )/\omega ^{2}$
as a function of the frequency $\omega $ for a system containing a
concentration of 1.5\% of defects with six weak springs (i.e. a total
concentration of 9\%) all having $K_{1}/K_{0}=0.1$. The continuos black line
shows $g(\omega )/\omega ^{2}$ for the system without defects, the red
dashed line the density calculated using equation (12) and the blue points
(connected by a line to guide the eye) that obtained from the simulations.

Fig 2. (Color on line) The change in the accumulated density of states $%
\Delta N(\omega )$ as a function of the frequency $\omega $ for a system
containing a concentration of 1.5\% of defects with six weak springs (i.e. a
total concentration of 9\%) all having $K_{1}/K_{0}=0.1$. The red continuous
line is the result of our calculations and the blue points are the results
of the simulations. Also shown is the calculated change $gtg(\omega )$
(dotted line) in the density of states for this system, so as to show the
positions of the low frequency peak and of the van-Hove singularities.

Fig 3. (Color on line) The reduced density of states $g(\omega )/\omega ^{2}$
as a function of the frequency $\omega $ for systems containing a 9\%
concentration of weak springs arranged in clusters of 6 (red dashed line,
marked 6), 4 (blue dotted line, marked 4) and 2 (olive solid line, marked 2)
bonds, for uncorrelated weak bonds (wine colored dashed line) and for the
system without any weak springs (black continuous line).

Fig 4. (Color on line) The change in the density of states, $gtg(\omega )$,
for systems containing a 9\% concentration of weak springs arranged in
clusters of 6 (red dashed line, marked 6), 4 (blue dotted line, marked 4)
and 2 (olive solid line, marked 2) bonds and for uncorrelated weak bonds
(wine colored dashed line). Figure 4a shows the full results, and figure 4b
an enlargement of the region with negative $gtg(\omega )$ in order to show
the source of the states contributing to the low frequency peak.

Fig.5 (Color on line) The density of states $g(\omega )$ as a function of
the frequency $\omega $ for a system containing a concentration of 1.5\% of
defects with six weak springs (i.e. a total concentration of 9\%) all having 
$K_{1}/K_{0}=0.1$.(red dashed line, marked 0.1), $K_{1}/K_{0}=0.2$.(blue
dotted line, marked 0.2), $K_{1}/K_{0}=0.3$.(olive short dash line, marked
0.3) and for the system with no weak springs (continuous black line)

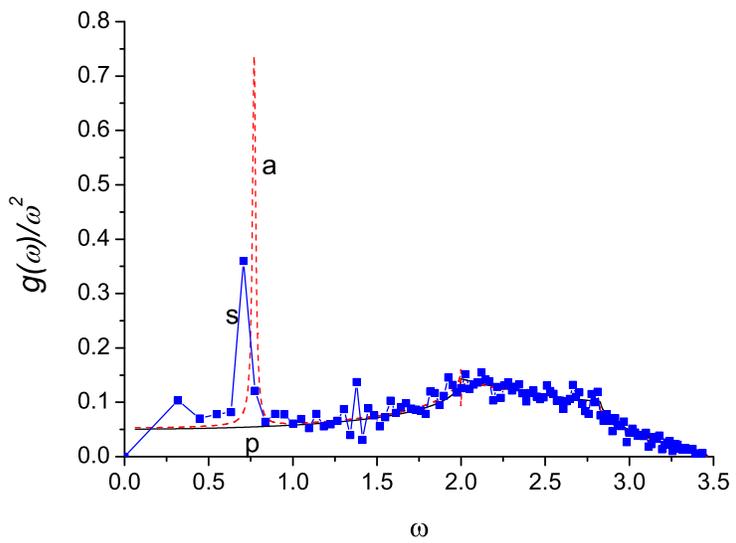
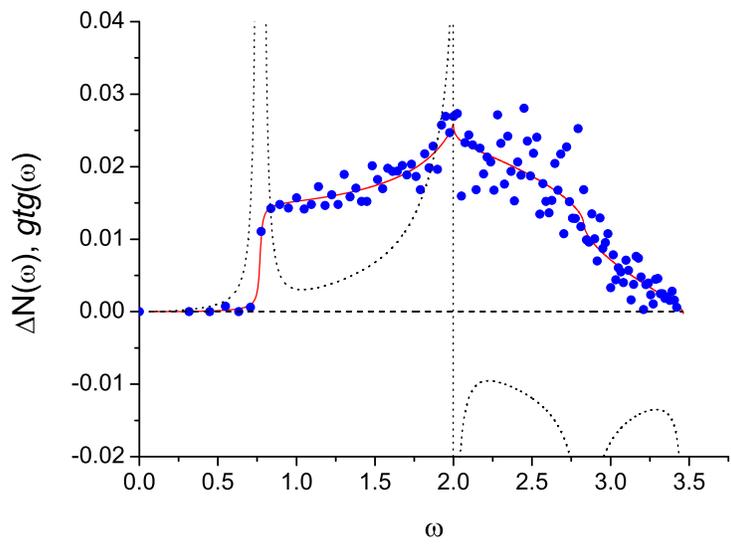
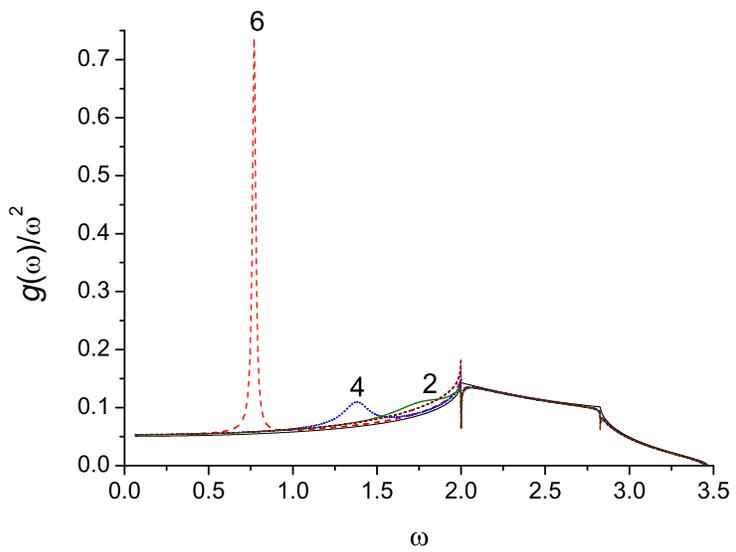
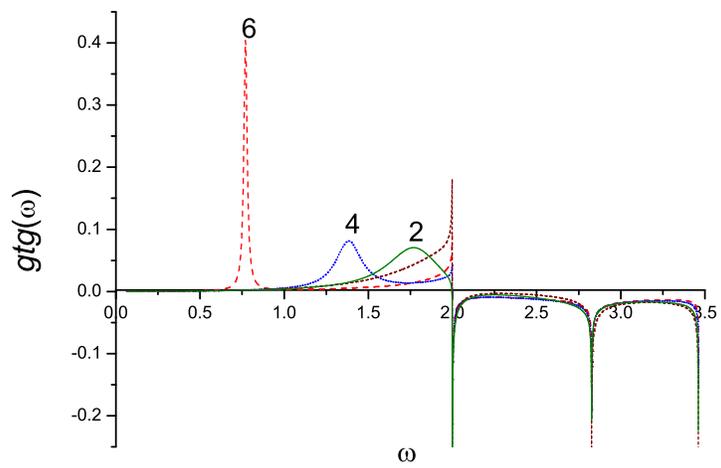
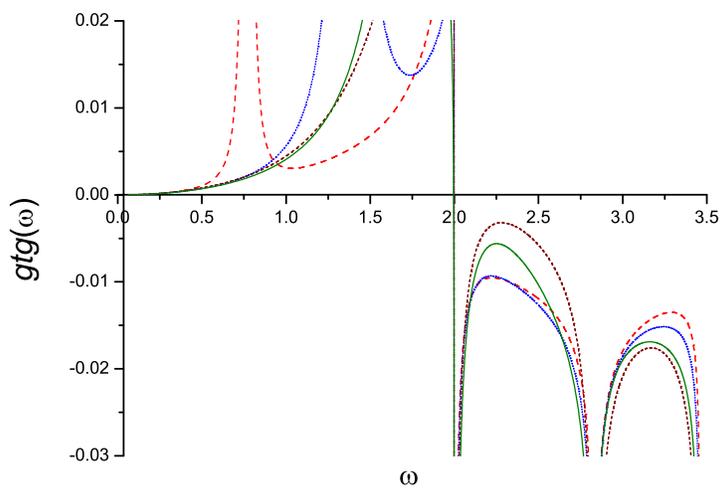
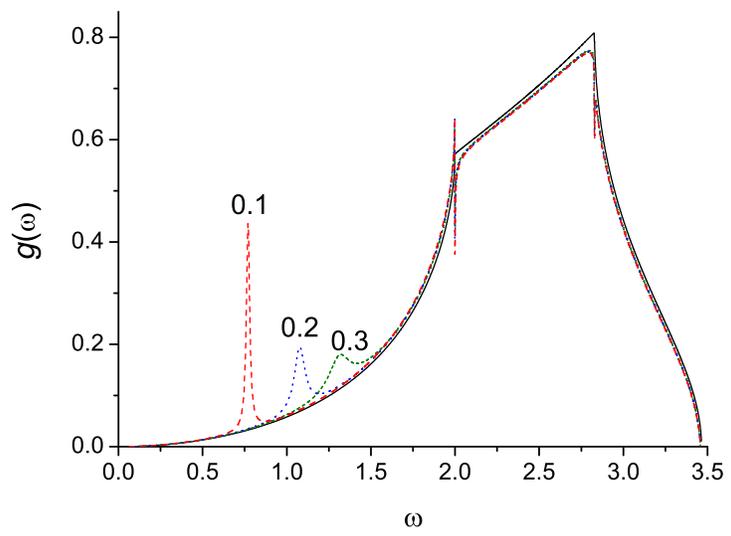
\end{document}